\documentstyle[11pt,aaspp4]{article}

\input psfig
\psfull

\parskip=\the\medskipamount
\def\etal{{\it et al. \/}}

\newcommand {\lsim}{\mbox{$\:\stackrel{<}{_{\sim}}\:$} }
\def\be{\begin{equation}}
\def\ee{\end{equation}}
\def\bea{\begin{eqnarray}}
\def\eea{\end{eqnarray}}
\begin{document}
\title{The Observational Case for Jupiter\\ 
Being a Typical Massive Planet}
\medskip

\author{Charles H. Lineweaver \& Daniel Grether \\
University of New South Wales\\
charley@bat.phys.unsw.edu.au}   

\begin{abstract}
We identify a subsample of the recently detected extrasolar planets 
that is minimally affected by the selection effects of the Doppler detection method.
With a simple analysis we quantify trends in the surface density of this 
subsample in the period-$M\sin (i)$ plane.
A modest extrapolation of these trends puts Jupiter in the most 
densely occupied region of this parameter space,
thus indicating that Jupiter is a typical massive planet rather
than an outlier.
Our analysis suggests that Jupiter is more typical than indicated by 
previous analyses. For example, instead of $M_{\rm Jup}$ mass exoplanets 
being twice as common as  $2\; M_{\rm Jup}$ exoplanets,
we find they are three times as common.
\end{abstract}

\section{Exoplanets and the Standard Model of Planet Formation}

The prevalence of infrared emission from accretion disks around 
young stars is consistent with the idea that such disks are ubiquitous. 
Their disappearance on a time scale of 50 to 100 million years 
suggests that the dust and gas accrete into planetesimals and 
eventually planets (Haisch \etal 2001). Such observations support 
the widely accepted idea that planet formation is a common by-product 
of star formation (e.g. Beckwith  \etal 2000).
In the standard model of planet formation, Earth-like planets accrete 
near the host star from rocky debris depleted of volatile elements, 
while giant gaseous planets accrete in the ice zones ($\gtrsim 4$ AU) 
around rocky cores (Lissauer 1995, Boss 1995).
When the rocky cores in the ice zones reach a critical mass 
($\sim 10 \: m_{\rm Earth}$) runaway gaseous accretion
(formation of jupiters) begins and continues until gaps form in the 
protoplanetary disk or the disk dissipates 
(Papaloizou \& Terquem 1999, Habing \etal 1999), leaving one or more
Jupiter-like planets at $\sim 4 - 10$ AU.

We cannot yet determine how generic the pattern described above is.
However, formation of terrestrial planets is thought to be less 
problematic than
the formation of Jupiter-like planets (Wetherill 1995).
Gas in circumstellar disks around young stars is lost within a few 
million years and it is not obvious that the rocky cores necessary to 
accrete the gas into a Jupiter can form on that time scale 
(Zuckerman \etal 1995). Thus, Jupiter-like planets may be rare.
Planets may not form at all if erosion, rather than growth, occurs 
during collisions of planetesimals (Kortenkamp \& Wetherill 2000).
The present day asteroid belt may be an example of such non-growth.
In addition, not all circumstellar disks produce an extant planetary system.
Some fraction may spawn a transitory system only to be accreted by the 
central star along with the disk (Ward 1997).
Also, observations of star-forming regions indicate that massive stars 
disrupt the protoplanetary disks around neighboring lower mass stars, 
aborting their efforts to produce planets (Henney \& O'Dell 1999).
Given these uncertainties,
whether planetary systems like our Solar System are common around Sun-like
stars and whether Jupiter-like planets are typical of such planetary
systems, are important open questions.

The frequency of Jupiter-like planets may also have implications 
for the frequency of life in the Universe. 
A Jupiter-like planet shields 
inner planets from an otherwise much heavier bombardment by 
planetesimals, comets and asteroids during the first billion years
after formation of the central star.
Wetherill (1994) has estimated that Jupiter significantly reduced 
the frequency of sterilizing impacts on the early Earth during the important
epoch $\sim 4$ billion years ago when life originated on Earth.
The removal of comet Shoemaker-Levy by Jupiter in 1994, is a more recent 
example of Jupiter's protective role.

To date (November, 2001), $74 $ giant planets ($M\sin (i) < 13\; M_{\rm Jup}$)
in close orbits ($< 4\; AU$) around  66 nearby stars have been detected 
by measuring the Doppler reflex of the host star (Marcy \etal 2001, Mayor
\etal 2001). Seven stars are host to multiple planets (six doubles, 
one triple system).
Approximately 5\% of the Sun-like stars surveyed 
possess such giant planets (Marcy \& Butler 2000).    
The large masses, small orbits, 
high eccentricities and high host metallicities of these 74 
exoplanets was not anticipated by theories of planet formation 
that were largely based on the assumption that planetary systems 
are ubiquitous and our Solar System is typical (Lissauer 1995).

\begin{figure*}[!ht]
\centerline{\psfig{figure=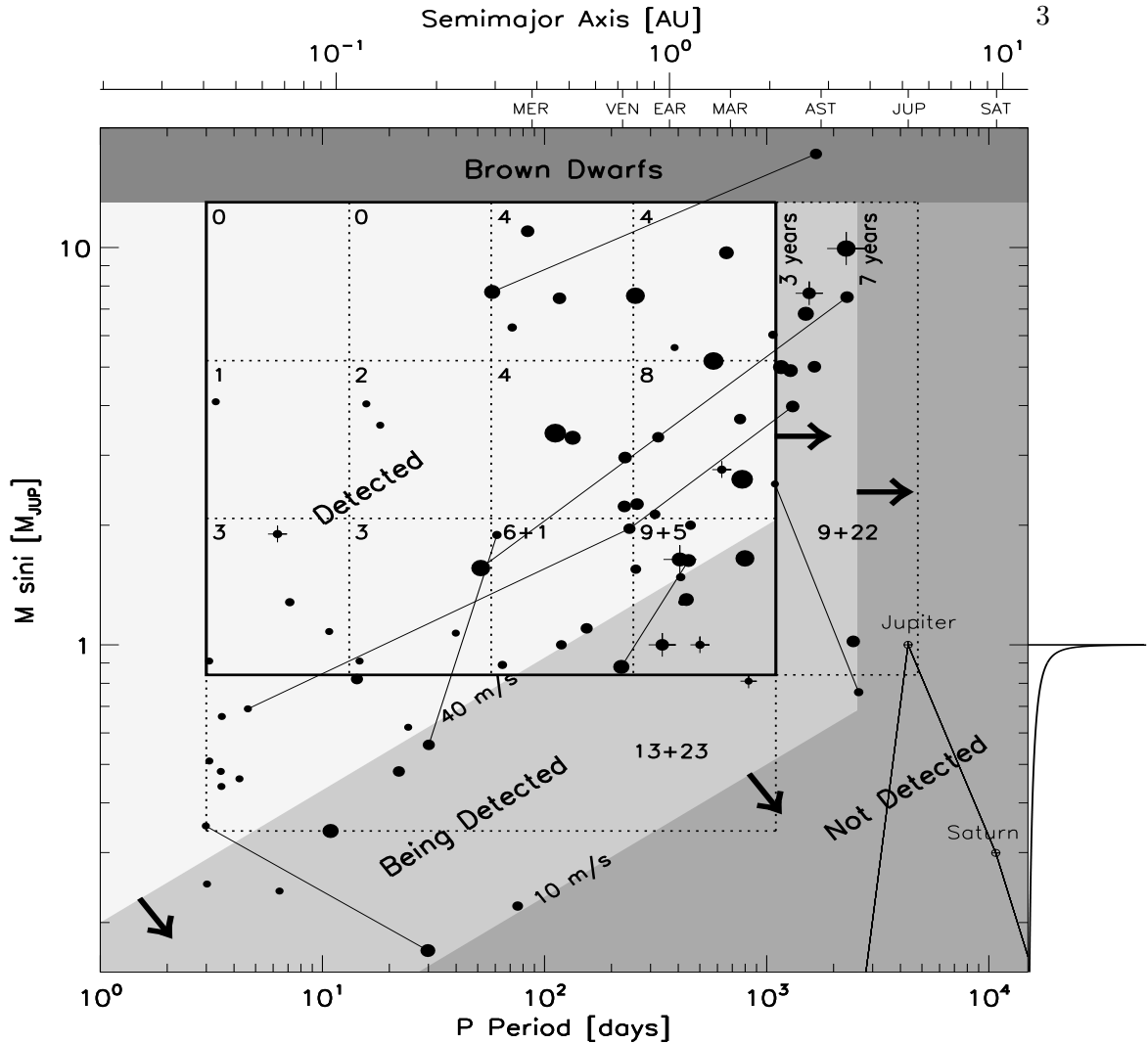,height=13.0cm,width=15cm}}
\caption{Mass as a function of period for the 74 exoplanets 
detected to date. Regions where planets are ``Detected'', 
``Being Detected'' and ``Not Detected'' by the Doppler surveys
 are shaded differently and represent the observational 
selection effects of the Doppler reflex 
technique (see Section \ref{sec:partitions} for a 
description of these regions).
The rectangle enclosing the grid of twelve boxes 
defines the subsample of $44$ planets less biased by selection effects. 
The numbers in the upper left of each box gives the number of planets in 
that box. The increasing numbers from left to right and from top to
bottom are easily identified trends. 
In Figs. \ref{fig:masshistogramfit} \& \ref{fig:periodhistogramfit}
we quantify and extrapolate these trends into the 
lower mass bin and into the longer period bin which includes Jupiter.
The ``+1'' and ``+5'' in the two boxes in the lower right
refer to the undersampling corrections discussed in Section \ref{sec:correction}.
The seven exoplanetary {\it systems} are connected by thin lines.
Jupiter and Saturn are in the ``Not Detected'' region.
The upper $x$ axis shows the distances and periods of the 
planets of our Solar System.
The brown dwarf region is defined by $M\sin (i)/M_{\rm Jup} > 13$. 
The point size of the exoplanets is proportional to the eccentricity 
of the planetary orbits. 
}
\label{fig:massperiodesp}
\end{figure*}

Naef \etal (2001) point out that none of the planetary companions 
detected so far resembles the giants of the Solar System. 
This observational fact however, is fully consistent with the idea that
our Solar System is a typical planetary system.
Fig. \ref{fig:massperiodesp} shows explicitly that selection effects 
can easily explain the lack of detections of Jupiter-like planets.
Exoplanets detected to date can not resemble the planets of our 
Solar System because the Doppler technique used to detect exoplanets
has not been sensitive enough to detect Jupiter-like planets.
If the Sun were a target star in one of the Doppler 
surveys, no planet would have been detected around it. 

This situation is about to change.
In the next few years Doppler planet searches will be making 
detections in the region of parameter space occupied by Jupiter.
Thus 
\clearpage
it is timely to use the current data to estimate how 
densely occupied that parameter space will be.
The detected exoplanets may well be the observable  $5\%$ tail of the 
main concentration of massive planets of which Jupiter is typical.
The main goal of this paper is to correct or account for
selection effects to the extent possible and then examine what the trends 
in the mass and period distributions indicate for the region
of parameter space near Jupiter.
Such an analysis is now possible because a statistically significant 
sample is starting to emerge from which we can determine meaningful 
distributions in mass, period as well as in eccentricity and metallicity. 
Our analysis helps answer the important question:
How does our planetary system compare to other planetary systems?

In the next section we present our method for identifying a less 
biased subsample of exoplanets. In Section \ref{sec:PMplane}
we identify and extrapolate the trends in mass and period.
In Section \ref{sec:discuss} we discuss our analysis and compare our 
results to previous work. In Section \ref{sec:summary} we summarize
our results.

\section{Period-Mass Plane}
\label{sec:PMplane}
\subsection{Selection Effects}
\label{sec:partitions}

Doppler surveys are responsible for all $74$ exoplanets plotted in
Fig. \ref{fig:massperiodesp}.
To detect an exoplanet, its host star must be Doppler-monitored 
regularly for a period $P_{\rm obs}$ greater than or comparable 
to the orbital period $P$ of the planet. Thus, one selection 
effect on the detection of exoplanets is
\be
P_{\rm obs} \gtrsim P.
\label{eq:P}
\ee
%
The relationship between the observable 
line-of-sight velocity of the host star, $K = v_{*}\sin (i)$, 
the mass of the planet, $M$, the mass of the host star, $M_{*}$, the 
velocity of the planet, $v_{\rm p}$, and 
the semi-major axis of the the planet's orbit, $a$, is obtained by 
combining $v_{\rm p} = (2\pi/P)\:\left[a/(1-e^{1/2})\right]$
(where $e$ is orbital eccentricity)
with momentum conservation, 
$M_{*}K = M\sin (i)\;v_{\rm p}$,  and  Kepler's third law 
$M_{*}= \frac{a^{3}}{P^{2}}$ (in the limits that $M_{*} >> M$, 
and where $M_{*}$, $a$ and $P$ are measured in 
solar masses, AU and years respectively).
Simultaneously solving these equations yields the induced line-of-sight velocity 
of the host star,
\be
K          = 2\pi \frac{M\sin (i)}{M_{*}^{2/3}} P^{-1/3}
              \left[\frac{1}{(1-e^{2})^{1/2}}\right].
\label{eq:K}
\ee

\noindent This equation is used to find $M\sin (i)$ as a function of
the Doppler survey observables $K, P$ and $e$, with $M_{*}$ estimated from
stellar spectra.
To detect an exoplanet, the radial velocity $K$ must be greater than 
the instrumental noise, $K_{\rm s}$. 
Thus, the Doppler technique is most sensitive to 
massive close-orbiting planets.
Fig. \ref{fig:massperiodesp} shows that we are now on the verge of 
being able to detect planetary systems like ours, i.e., 
Jupiter-mass planets at $\gtrsim 4$ AU from nearby host stars. 
The grey regions of Fig. \ref{fig:massperiodesp}
partition the parameter space and represent the 
selection effects of the Doppler surveys.
We use these partitions to identify a 
less biased subsample of 44 exoplanets within the rectangular area
enclosed by the thick solid line.

The largest observed $P$ and the smallest observed $K$ of the exoplanets 
in Fig. \ref{fig:massperiodesp} are inserted into Equations 
\ref{eq:P} \& \ref{eq:K} to empirically define
the boundary between the ``Being Detected'' and the ``Not Detected'' regions 
in Fig. \ref{fig:massperiodesp}. 
To define the ``Detected'' region of parameter space in which virtually 
all planets should have been detected (thus defining a less biased 
subsample of exoplanets)
we consider planets with $P < 3$ years with $K > 40$ m/s, 
that have been observed for more than 3 years with an instrumental
noise $K_{\rm s} < 20$ m/s. 
The rectangle in Fig. \ref{fig:massperiodesp} is 
the largest rectanglar area that approximately fits inside the 
``Detected'' region.
This method of cutting the data to remove biases is reminiscent of 
the astronomical practice of constructing a volume-limited sample 
from a magnitude-limited sample.  The area within the rectangle subsumes the ranges
$3 < P < 1000$ days and $0.84 < M\sin (i)/M_{\rm Jup} < 13$ and is
subdivided into a minimum number of 
smaller areas (12 boxes) for histogram binning convenience
(Figs. \ref{fig:masshistogramfit} \& \ref{fig:periodhistogramfit}). 
Trends in $M \sin (i)$ and $P$ identified within this subsample are 
less biased than trends based on the full sample of exoplanets.

If Jupiter is a typical giant planet, the region around Jupiter
in the $P - M\sin (i)$ plane of Fig. \ref{fig:massperiodesp}
will be more densely occupied than other regions --
the density of planets in the lower right will
be larger than in the upper left. Although we are dealing with small 
number statistics, that trend is the main identifiable trend
in Fig. \ref{fig:massperiodesp}; the number of exoplanets in
the boxes increases from left to right and top to bottom.
In the rest of the paper we quantify and extrapolate these trends 
into the lower mass bin and longer period bin enclosed by the dotted 
rectangles in Fig.
\ref{fig:massperiodesp}.

\subsection{Undersampling Corrections}
\label{sec:correction}

\begin{figure*}[!ht]
\centerline{\psfig{figure=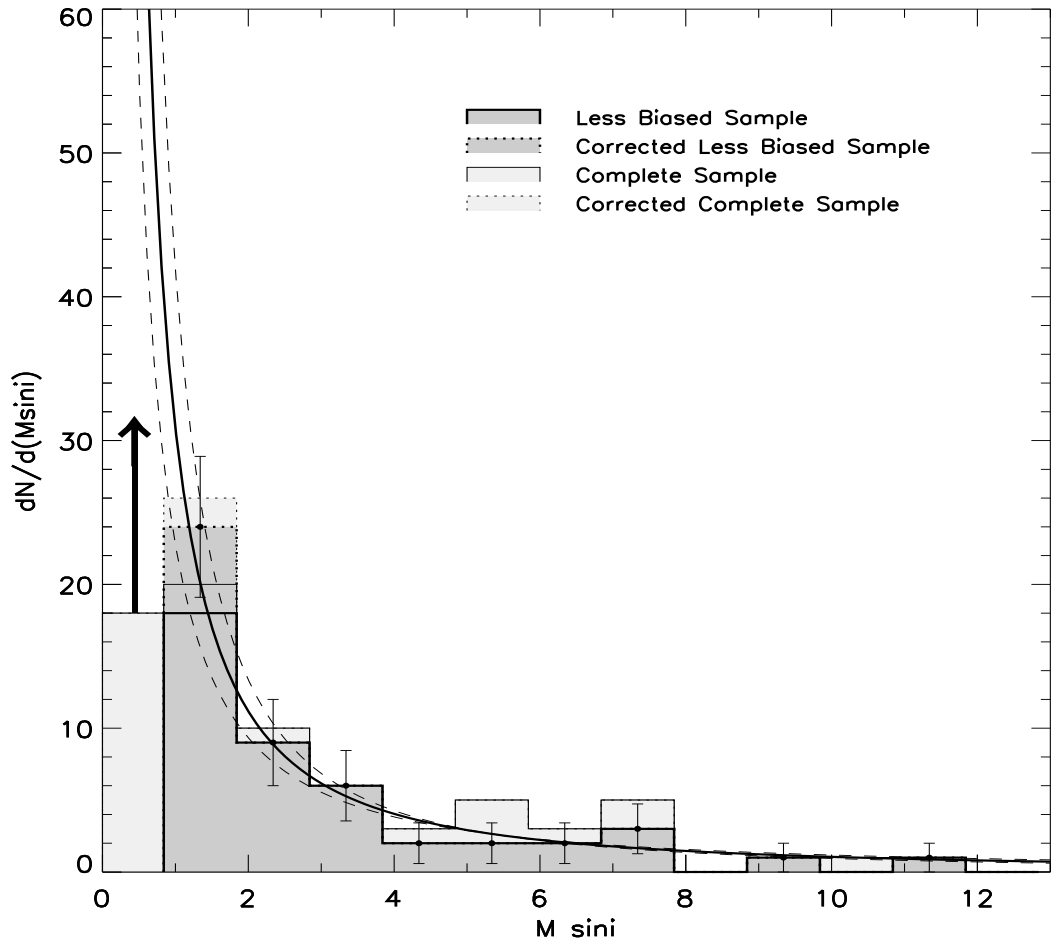,height=13.0cm,width=15cm}}
\caption{Mass histogram of the less-biased subsample (dark grey) of 
44 exoplanets within the rectangle in 
Fig. \ref{fig:massperiodesp}  compared to the histogram of
the complete sample of 74 exoplanets (light grey).
The errors on the bin heights are Poissonian.
The solid curve and the enclosing dashed curves are the best fit 
and $68\%$ confidence levels from fitting the functional form 
$(dN/dM\sin i) \propto (M\sin i)^{\alpha}$ to the histogram of 
the corrected less-biased subsample (50 = 44 + 6 exoplanets).
The extrapolation of this curve into the lower 
mass bin produces an estimate of 
the substantial incompleteness of this bin (arrow).
The mass ranges of the bins are: $0.84, 1.84, 2.84... M\sin (i)/M_{\rm Jup}$. 
The lower 
limit of $0.84$ was chosen to match the lower limit of the logarithmic
binning in Fig. \ref{fig:massperiodesp}.}
\label{fig:masshistogramesp}
\end{figure*}

\begin{figure*}[!ht]
\centerline{\psfig{figure=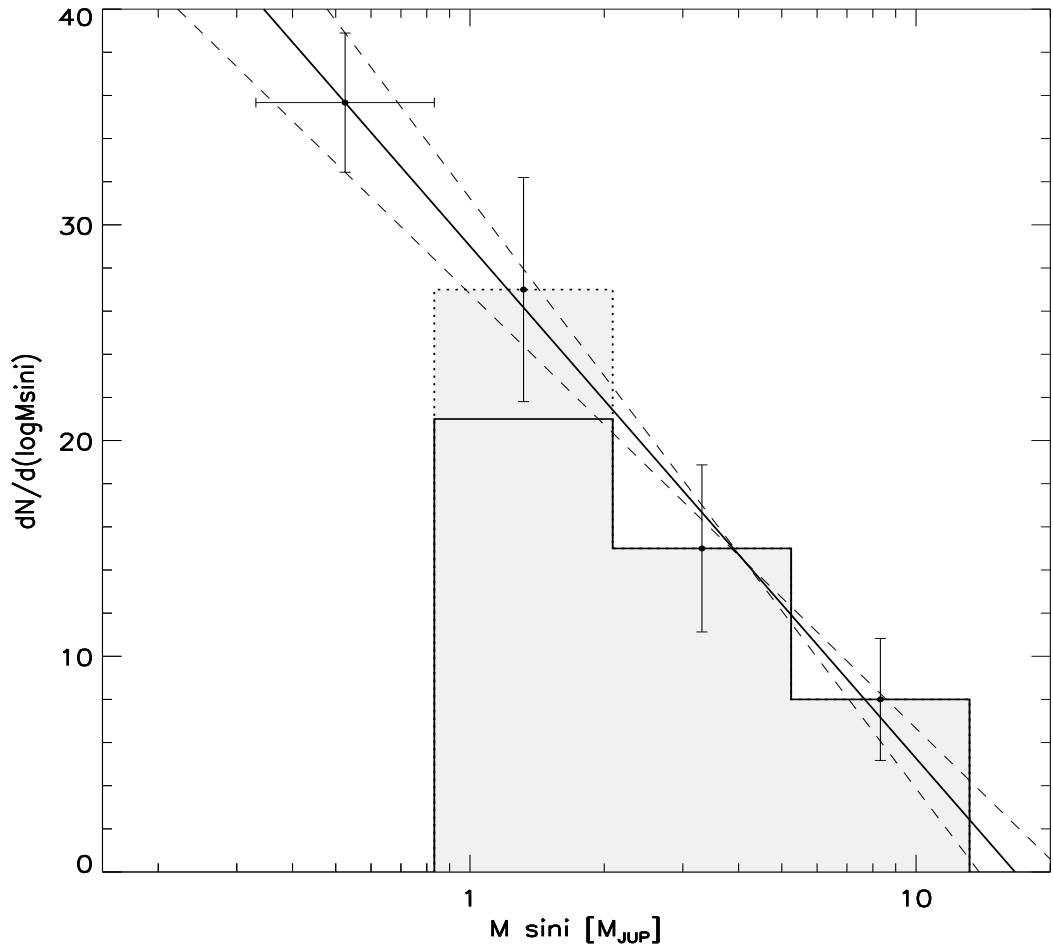,height=13.0cm,width=15cm}}
\caption{Histogram in log($M \sin i$) of the 50 (= 44 + 6) exoplanets 
within the rectangular area enclosed by the thick solid line
in  Fig \ref{fig:massperiodesp}.
The difference between the solid and dotted histograms in the 
lowest mass bin is the correction factor due to undersampling
as described in Section \ref{sec:correction}.
The line is the best fit to the functional form 
$dN/d(log(M\sin i)) = a\; log(M\sin i) + b$. 
The best fit slope is $a = -24 \pm 4$.
The extrapolation of this trend into the adjacent lower mass bin 
($0.34 < M\sin (i)/M_{\rm Jup} < 0.84$) indicates  that at least 
$\sim 36 \pm 3$ exoplanets with periods
in the range $3 - 1000$ days  are being hosted by the target stars now 
being monitored. This is 23 more than the 13 that have been detected
to date in this mass range.
}
\label{fig:masshistogramfit}
\end{figure*}

\begin{figure*}[!ht]
\centerline{\psfig{figure=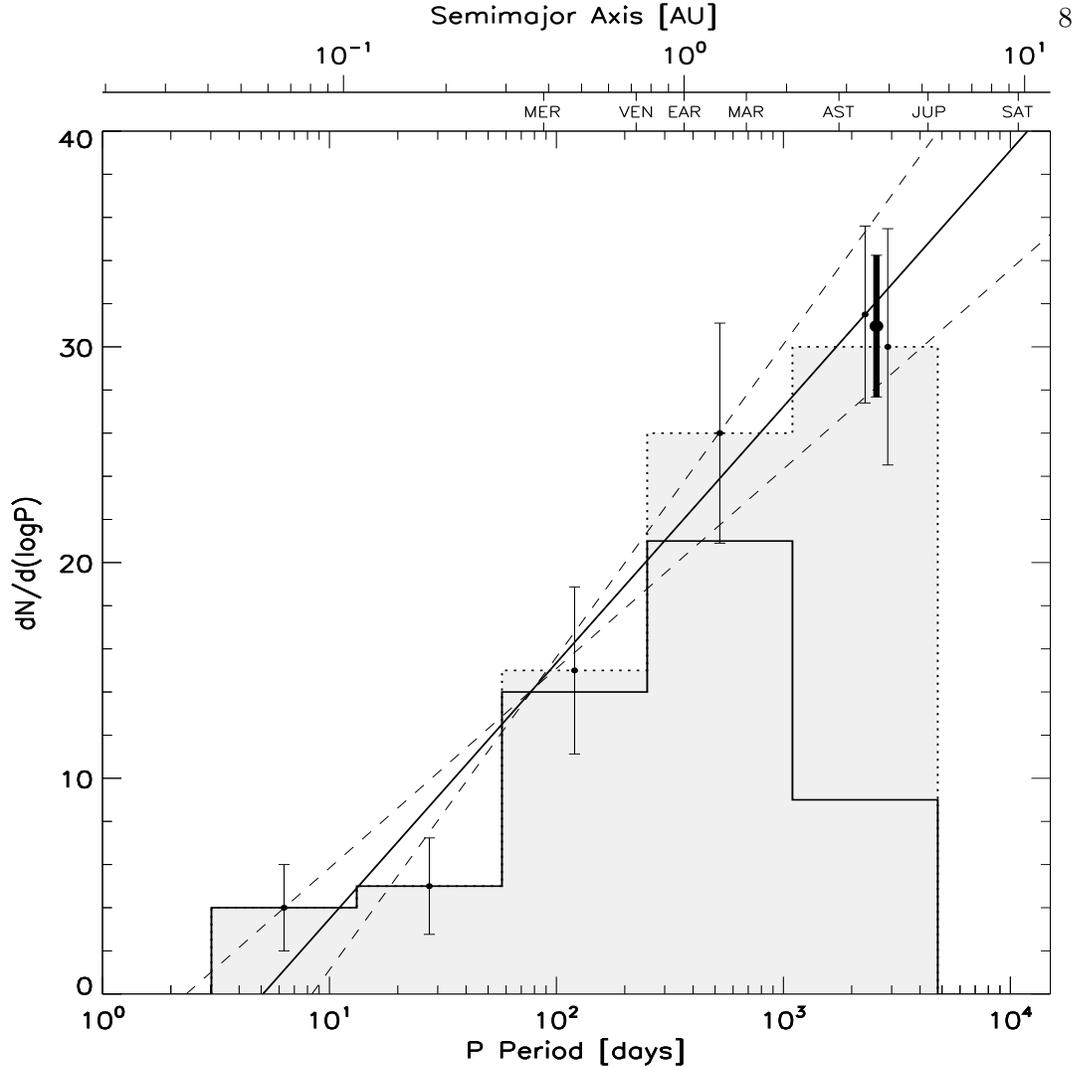,height=13.0cm,width=15cm}}
\caption{Trend in period of the corrected (dotted) and uncorrected
(solid) less-biased subsample.
The line is the best fit to the corrected histogram.
The functional form fitted is linear in $log P$, i.e.,
$dN/d(log P) = a\; log P + b$. The best fit slope is $a = 12 \pm 3$.  
We estimate the number in the longest period bin
($1000 < P < 5000$ days) in two independent ways:
1) based on the extrapolation of the linear fit to this longer period bin
and 2) correcting for undersampling in the ``Being Detected'' region 
as described in Section \ref{sec:correction}.
The former yields $32 \pm 4$ while the later yields $30 \pm 6$. We take 
the weighted mean of these, $31  \pm 3$, as our best prediction for 
how many planets will be found in this longest period bin scattered over 
the mass range $0.84 < M\sin (i)/M_{\rm Jup} < 13$.
To date, 9 extrasolar planets have been found in this period bin.
Thus we predict that $22 \pm 3$ more planets will be found in this period bin.
}
\label{fig:periodhistogramfit}
\end{figure*}

Within the rectangle enclosed by the thick solid line in 
Fig. \ref{fig:massperiodesp}, two boxes in
the lower right lie partially in the ``Being Detected'' region.
Thus they are partially undersampled compared to the other boxes within
the rectangle.
We correct for this undersampling by making the simple assumption that the 
detection efficiency is linear in the ``Being Detected'' region.  
That is, we assume that the detection efficiency is $100\%$ in 
the ``Detected'' region and $0\%$ in the ``Not Detected'' region 
and decreases linearly inbetween.
This linear correction produces the ``+1'' and ``+5'' corrections to 
the number of exoplanets observed in these two boxes and produces
the dotted corrections to the histograms in Figs. \ref{fig:masshistogramesp},
\ref{fig:masshistogramfit} and \ref{fig:periodhistogramfit}.
\clearpage
\vspace{4cm}

\clearpage

\clearpage

Eight new planets were detected while this paper was in preparation 
(Vogt \etal 2002, Tinney \etal 2002). These are distinguished in 
Fig. \ref{fig:massperiodesp} by crosses plotted over the dots. 
Five of the eight justify our parameter space partitions by falling 
as expected, in the ``Being Detected'' region. 
The other three fall unexpectedly in the ``Detected'' 
region. If this region were fully detected newly detected planets 
would not fall there. However, two of the three 
(HD 68988 \& HD 4203, both with $M\sin (i)$ between 1.5 and 2 $M_{\rm Jup}$) 
had only been observed for 1.5 years and therefore do not  
qualify for our least biased sample containing only host stars 
that have been monitored for at least three 
years. Therefore, these two apparent anomalies do not undermine our 
parameter space partitions.
The third host lying in the ``Detected'' region lies near the 
``Being Detected'' boundary. It has a two year period and was monitored 
for four years at the Anglo-Australian Telescope (AAT) (Tinney \etal 2002). 
Its observed velocity of $\sim 50$ m/s made it a $\sim 15 \sigma$ signal 
with the AAT's $\sim 3 $ m/s sensitivity.
Its appearance in the ``Detected'' region may be ascribable to 
late reporting or may reflect the need to combine the constraints from
Eqs. \ref{eq:P} \& \ref{eq:K} into a smooth curve rather than 
two straight boundaries.
Subsequent exoplanets can be similarly used to verify the accuracy of 
our representation of the Doppler detection selection effects in the $P - M \sin (i)$ plane.

\subsection{Mass and Period Histogram Fits}
\label{sec:fits}

The distribution of the masses of the exoplanets is 
shown against $M\sin (i)$ in Fig. \ref{fig:masshistogramesp} 
and against log $M\sin (i)$ in Fig. \ref{fig:masshistogramfit}. 
The 6 planet correction (Section \ref{sec:correction})
to the $M\sin (i) \sim 1$ bin in both figures is indicated by the 
dotted lines.
In Fig. \ref{fig:masshistogramesp} the solid curve and the 
enclosing dashed curves are the best fit and $68\%$ confidence 
levels of the functional form 
$(dN/dM\sin i) \propto (M\sin i)^{\alpha}$ fit to the histogram of 
the corrected less-biased subsample (50 = 44 + 6 exoplanets).
We find $\alpha = -1.5 \pm 0.2$.
This means, for example, that within the same period range there are 
$\sim 3$ times as many $M_{\rm Jup}$ as $2\;M_{\rm Jup}$ exoplanets and
similarly $\sim 3$ times as many $0.5\; M_{\rm Jup}$  as $M_{\rm Jup}$ exoplanets.
This slope is steeper than the $\alpha \approx -1.0$ of previous
analyses (Section \ref{sec:compare}).

In Fig. \ref{fig:masshistogramfit} the line is the best fit of 
the functional form 
$dN/d(log(M\sin i)) = a\; log(M\sin i) + b$
to the histogram. 
The best-fit slope, $a = -24 \pm 4$, is significantly 
steeper than flat. 
The extrapolation of this trend into the adjacent lower mass bin 
($0.34 < M\sin (i)/M_{\rm Jup} < 0.84$) indicates  that at least 
$\sim 36 \pm 3$ exoplanets with periods
in the range $3 - 1000$ days  are being hosted by the target stars now 
being monitored. Since 13 have been detected
to date in this mass range, we predict that $23 \pm 3$ more have yet to be detected.
Thus we predict that the continued monitoring of the target stars
that produced the current set of exoplanets will eventually yield
$\sim 23$ new planets in the parameter range
$ 3 < P < 1000$ days with $0.34 < M\sin (i)/M_{\rm Jup} < 0.84$ 
(dotted horizontal rectangle in Fig. \ref{fig:massperiodesp}). 

The trend of increasing number of exoplanets per box as one descends in
mass does not hold true in the highly undersampled longest period bin
($1000 < P < 5000$). The absence of this trend may be 
the result of small number statistics or an 
additional 
\begin{figure*}[!ht]
\centerline{\psfig{figure=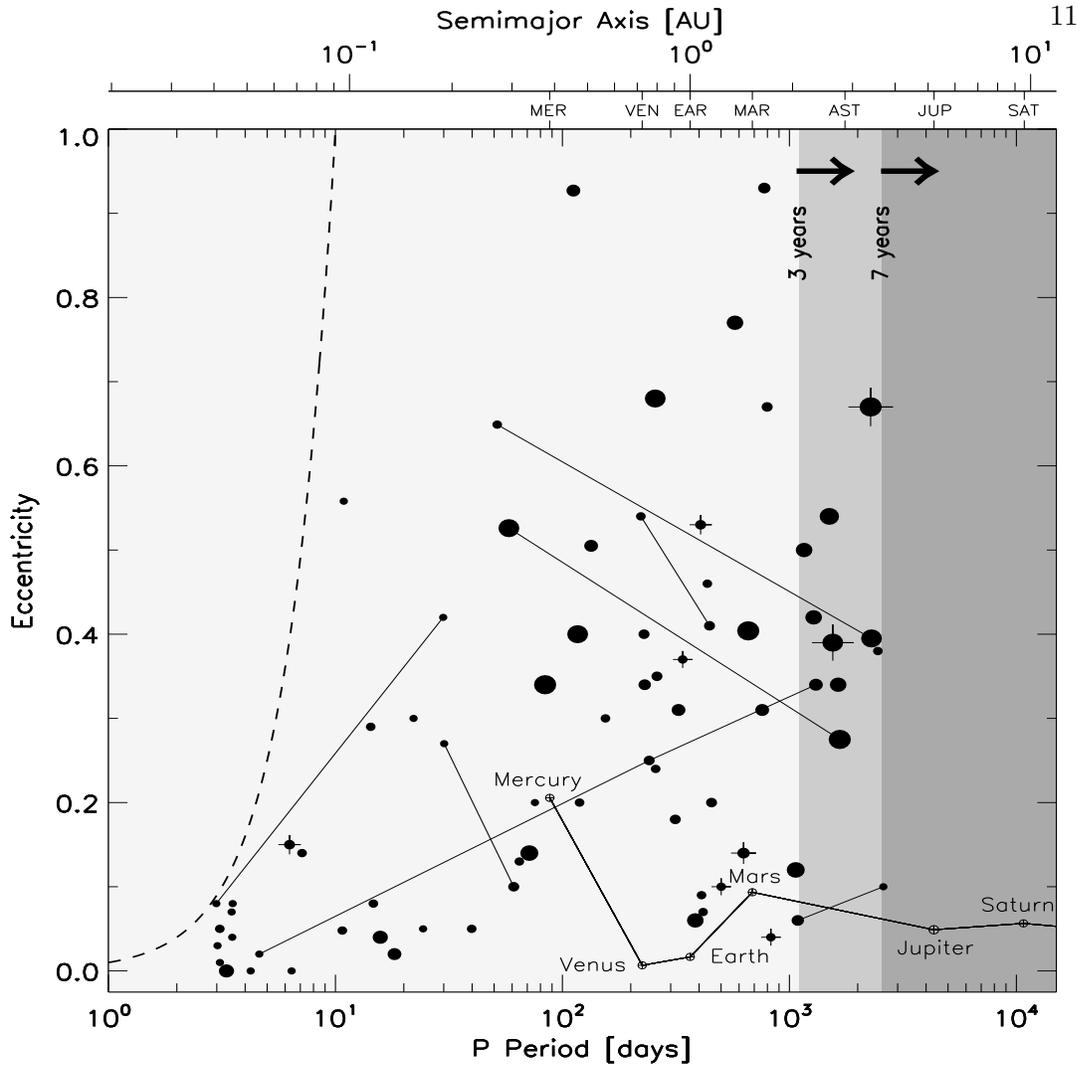,height=13.0cm,width=15cm}}
\caption{Eccentricity of the orbits of exoplanets as a function of period.
Planets in the same system are connected by lines.
Notice that in four out of the seven planetary systems the more 
distant member is less eccentric and more massive. This is 
what one would expect if the Solar System is a typical planetary system
and  Doppler-detectable exoplanets have been scattered in by more
massive, less eccentric companions.
This also suggests that the exoplanets closer to Jupiter's region of 
$P - M\sin (i)$ parameter space may share Jupiter's low eccentricity.
In two of the planetary systems that do not conform to this pattern,
the inner planet is so close to the central star that its orbit has 
probably been tidally circularized -- an $r^{-3}$ effect indicated 
by the dashed line.
In the one remaining exception, the eccentricities are low and
do resemble the planets of our Solar System.
Thus, in  planetary systems, the eccentricities of exoplanets with 
longer periods (like Jupiter's) tend to also have less eccentric 
orbits (like Jupiter's).
Both the high metallicity of exoplanet hosts and this trend in 
eccentricity lend some support to the gravitational scattering
model for the origin of hot jupiters, suggesting that Jupiter
is a fairly typical massive planet of the more typical lower 
metallicity systems.
As more exoplanetary systems are detected this pattern can be readily tested.
Pointsize is proportional to exoplanet mass.
}
\label{fig:eccperiodesp}
\end{figure*}

\noindent hint that a smooth curve, rather than our two straight 
boundaries, more accurately describes the selection effects.

Extrapolation of the linear trend found in Fig. \ref{fig:periodhistogramfit}
indicates that 22 new planets will be discovered in the first bin
to the right of the rectangle in Fig. \ref{fig:massperiodesp} 
($ 1000 < P < 5000$ days in the mass range $0.84 < M\sin (i)/M_{\rm Jup} < 13$). 
Following the trend in $M\sin (i)$ identified in 
Fig. \ref{fig:masshistogramfit}, these 22 should be preferencially 
assigned to the lower masses in this range.
Extrapolation of the trend in period into an even longer period bin, which 
would include Saturn, is more problematic.

\section{Discussion}
\label{sec:discuss}
\subsection{Eccentricity}

A significant difference between the detected exoplanets
and Jupiter, is the high orbital 
eccentricities of the exoplanets.
The eccentricities of the planets of our Solar System
were presumably constrained to small values ($e \lsim 0.1$) 
by the migration through, and accretion of,
essentially zero eccentricity disk material.
A simple model that can explain the higher exoplanet eccentricities
is that in higher 
metallicity systems, the higher abundance of refractory material in 
the protoplanetary disk may lead to the production of more planetary 
cores in the ice zone producing multiple Jupiters which 
gravitationally scatter off each other. Occasionnaly one will be scattered
in closer to the central star and become Doppler-detectable 
(Weidenschilling \& Marzari 1996).
%
If that is the origin of the hot jupiters, then 
the detected exoplanets may be the high metallicity tail of a 
distribution in which our Solar System is typical, and
as longer period giant 
planets are found they will have 
lower eccentricities, 
comparable to Jupiter's and Saturn's. 
Thus, if Jupiter is the norm rather than 
the exception, not only will we find more planets in the $P -M\sin (i)$ 
parameter space near Jupiter as reported above, but also the 
eccentricities of the longer period exoplanets will be lower.
The general distribution of the eccentricities of the exoplanets does 
not seem to reflect this, but exoplanets in planetary systems 
lend some support to the idea (see Fig. \ref{fig:eccperiodesp} and caption).

\clearpage
\begin{figure*}[!ht]
\centerline{\psfig{figure=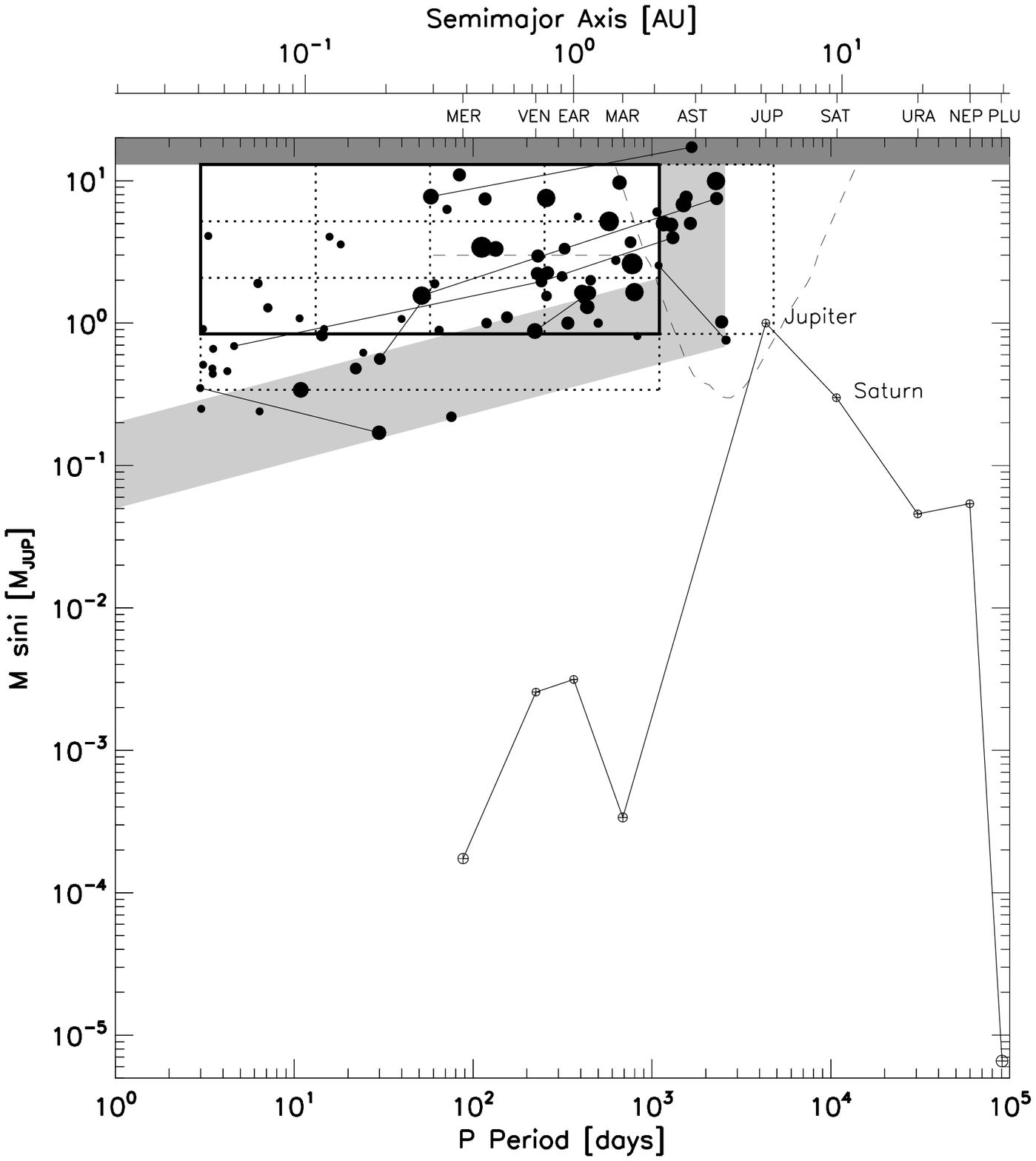,height=13.0cm,width=15cm}}
\caption{The region of the $P - M\sin (i)$
plane occupied by our Solar System compared to
the region being sampled by Doppler surveys.
Doppler surveys are on the verge of detecting
Jupiter-like exoplanets. The lines, symbols and shading are the 
same as in Fig. \ref{fig:massperiodesp}.
We would like to know how planetary systems in general
are distributed in this plane.
Extrapolations of the trends quantified in this paper put 
Jupiter in the most densely occupied region of the 
$P - M\sin (i)$ parameter
space indicating that the detected exoplanets are the observable 
tail of the main concentration of massive planets 
that occupies the parameter space closer to Jupiter.
If our Solar System is typical then the dispersion away from Jupiter 
into the Doppler-detectable region of this  plot may be
largely due to the effect of high metallicity in producing
gravitational scattering.  
Whether Jupiter is slightly more or less massive than the average most
massive planet in a planetary system is difficult to determine.
However, the strong correlation between the presence of Doppler-detectable
exoplanets and high host metallicity (e.g. Lineweaver 2001) suggests that
high metallicity systems preferentially produce massive Doppler-detectable
exoplanets. This further suggests (since the Sun is more metal-rich 
than $\sim 2/3$ of local solar analogues) that Jupiter may be 
slightly more massive than the average most massive planet of 
an average metallicity, but otherwise Sun-like, star.
The dashed wedge-shaped contour represents the microlensing constraints discussed in the text.
}
\label{fig:massperiodmore}
\end{figure*}

\subsection{Fitting for $\alpha$}
\label{sec:compare}

In this paper we have focused on 
the position of Jupiter relative to the exoplanets. 
This relative comparison does not require a conversion of 
exoplanet $M\sin (i)$ values to $M$ values 
(Jorissen \etal 2001, Zucker \& Mazeh 2001a, 2001b and 
Tabachnik \& Tremaine 2001).
However, for this comparison Jupiter's position needs to be lowered and spread
out a bit.
Given a random distribution of orbital inclinations, the 
probability of $y =M \sin (i)$, given $M$, is
\be
P(y|M) = \frac{y}{M^{2}\sqrt{1-\frac{y^{2}}{M^{2}}}}.
\ee
With $M = M_{\rm Jup}$, this probability is the curve placed outside 
the plotting area on the 
lower right of Fig. \ref{fig:massperiodesp}. 
It represents the region of $M\sin (i)$ that Jupiter-mass planets 
would fall in when observed at random orientations.
The mean of this distribution is $\pi/4 \approx 0.79$ while 
the median is $0.87$ (in units of $M_{\rm Jup}$).
This lowers and spreads out in $M\sin (i)$ the position of Jupiter
but does not change the main results of the extrapolations done here.

The functional form $dN/d(M\sin i) \propto (M\sin i)^{\alpha}$
can be fit in various ways to various versions of the $M\sin (i)$ histogram of
exoplanets. When the histogram of all 74 exoplanets is fit, including 
the highly undersampled lowest $M\sin (i)$ bin, the result is 
$\alpha = -0.8$. This is reported in Marcy \etal 2001
and we confirm this result.
This value for $\alpha$ is  close to the $\approx -0.8 \pm 0.2 $ found 
for very low mass stars (Bejar \etal 2001).
When the lowest exoplanet $M\sin (i)$ bin is ignored because of known
incompleteness we obtain $\alpha = -1.1$.
This is very similar to the  $\alpha \approx -1.1$ found in fits to the
derived $M$ distribution (Zucker \& Mazeh 2001a, Tabachnik \& Tremaine 2001).
The fit for $\alpha$ seems to be more dependent on how the first bin 
is treated and how the sample for fitting is selected 
than on whether one fits to $M\sin (i)$ or $M$.

Fitting to the $M\sin (i)$ histogram of the less biased sample of 
44 exoplanets, uncorrected for undersampling, yields $\alpha = -1.3$. 
After correcting for undersampling as described in 
Section \ref{sec:correction} we obtain our final result: 
$\alpha = -1.5 \pm 0.2$ (Fig. \ref{fig:masshistogramesp}).
This slope is steeper than the $\alpha \approx -1.0$ of previous
analyses and indicates that instead of $M_{\rm Jup}$ mass exoplanets 
being twice as common as  $2\; M_{\rm Jup}$ exoplanets,
they are three times as common.

\section{Summary}
\label{sec:summary}
Despite the fact that massive planets are easier to detect, the 
mass distribution of detected planets is strongly peaked toward the 
lowest detectable masses.
And despite the fact that short period planets are easier to detect, the 
period distribution
is strongly peaked toward the longest detectable periods. 
To quantify these trends as accurately as
possible, we have identified a less-biased subsample of exoplanets 
(Fig. \ref{fig:massperiodesp}).
Within this subsample, we have identified trends in $M\sin (i)$ and period 
that are less biased than trends based on the full sample 
of exoplanets.
Straightforward extrapolations of the trends quantified here, into the area of 
parameter space occupied by Jupiter, indicates that Jupiter lies 
in a region of parameter space densely occupied by exoplanets.

Our analysis indicates that 45 new planets will be detected in 
the parameter space discussed in the text. This estimate of 45 is a lower
limit in the sense that  if a smooth curve, rather than
our two straight boundaries, more accurately describes the selection effects
in Fig \ref{fig:massperiodesp}, larger corrections to the bin numbers
would steepen the slopes in both 
Fig. \ref{fig:masshistogramfit} \&  \ref{fig:periodhistogramfit}.

Despite the importance of the mass distribution and
the trends in it, it is the trend in period
that, when extrapolated, takes us to Jupiter and 
the parameter space occupied by Jupiter-like exoplanets 
(compare Figs. \ref{fig:massperiodesp} \& \ref{fig:periodhistogramfit}).
Long term slopes in the velocity data that have not yet been associated with planets 
are present in  a large fraction of the target stars surveyed with the Doppler 
technique (Butler, Mayor private communication).
However, quantifying the percentage of host stars showing such residual trends is difficult 
and depends on instrumental noise, phase coverage and the signal to noise threshold used to decide 
whether there is, or is not, a long term trend.

Figure \ref{fig:massperiodmore} shows that the Doppler technique has been able to 
sample a very specific high mass, short-period region of the $log P - log M\sin (i)$ plane. 
Thus far, this sampled region does not overlap with the 10 times 
larger area of this plane occupied by the nine planets of our Solar System. 
Thus there is room in the $\sim 95\%$ of target
stars with no Doppler-detected planets, to harbour planetary systems 
like our Solar System.

The trends in the exoplanets detected thus far do not rule out the 
hypothesis that our Solar System is typical.
They support it.
The extrapolations of the trends quantified here 
put Jupiter in the most densely occupied region of the
$P - M\sin (i)$ parameter space. 
Our analysis suggests that Jupiter is
more typical than indicated by previous analyses -- instead of 
$M_{\rm Jup}$ mass exoplanets being twice as common as 
$2\; M_{\rm Jup}$ planets we find they are three times as common.
In addition long term trends in velocity, not yet identified 
with planets, are common.
Both of these observations indicate that the detected exoplanets 
are the observable tail of the main concentration of 
massive planets of which Jupiter is likely to be a typical member rather than an outlier.

Null results from microlensing searches have been used to constrain the frequency 
of Jupiter-mass planets (Gaudi et al. 2002). These are plotted in Fig. \ref{fig:massperiodmore}. 
Less than $33\%$ of the lensing objects (presumed to be Galactic bulge M-dwarfs) have planetary companions 
within the dashed wedge-shaped area (the period scale, but not the AU scale, is applicable to this area). 
A conversion of the relative frequencies reported here to a fractional abundance in the wedge-shaped area
yields the rough estimate that more than $\sim 3$ percent of Doppler-surveyed Sun-like stars will be found 
to have companions with masses and periods in the wedge-shaped area. 
Thus our results are crudely consistent with current microlensing constraints.
However, because of the difference in host mass, ($\sim M_{Sun}$ for Doppler surveys and $\sim 0.3 M_{Sun}$ for microlensing) 
it is not clear that such a direct comparison is meaningful.
For example, if in the next few years Doppler and microlensing constraints appear to conflict, it may simply be that 
typical planetary masses scale with the mass of the host star, that is, Jupiter-mass planets at Jupiter-like orbital radii
may be more common around $\sim M_{Sun}$ stars than around $\sim 0.3 M_{Sun}$ stars. 

\section{Acknowledgements}
We thank Penny Sackett, Scott Gaudi, Ross Taylor and an anonymous referee for helpful suggestions.
CHL is supported by an Australian Research Council research fellowship.

\clearpage
\section{References}

\noindent Beckwith, S.V.W., Henning, T. \& Nakagawa, Y 2000,
``Dust Properties and Assembly of Large  Particles in Protoplanetary
Disks'' In ``Protostars and Planets IV'' edt V. Mannings, A.P. Boss
\& S.S. Russell, pp 533-558, Univ. Arizona Press, Tucson

\noindent Bejar, V.J.S. \etal 2001  `The Substellar Mass Function of Sigma Orionis' \apj, 556, 830-836

\noindent Boss, A.P. 1995 ``Proximity of Jupiter-like Planets to Low-mass Stars''
Science, 267, 360

\noindent Gaudi, B.S. \etal 2002, ``Microlensing Constraints on the Frequency of Jupiter-Mass
Companions: Analysis of Five Years of PLANET Photometry'', \apj,  566, 463

\noindent Habing, H.J. \etal 1999, ``Disappearance of stellar debris disks around main
sequence stars after 400 million years'' Nature, 401, 456-458

\noindent Haisch, K.E. \etal 2001, `Disk Frequencies and Lifetimes of Young Clusters' \apj, 553, L153-156

\noindent Henney, W.J. \& O'Dell, C.R. 1999, `A Keck High-Resolution Spectroscopic Study of the Orion
Nebula Proplyds', AJ, 118, 2350-2368

\noindent Jorissen, A., Mayor, M. \& Udry, S. 2001
``The distribution of exoplanet masses''  A\&A submitted, astro-ph/0105301

\noindent Kortenkamp \& Wetherill, G.W. 2000,`Terrestrial Planet and Asteroid Formation in the Presence
of Giant Planets' Icarus, 143, 60

\noindent Lineweaver, C.H. 2001, ``An Estimate of the Age Distribution of Terrestrial Planets in the
Universe: Quantifying Metallicity as a Selection Effect'' Icarus, 151, 307-313

\noindent Lissauer, J.J. 1995, ``Urey Prize Lecture: On the Diversity of 
Plausible Planetary Systems'' Icarus, 114, 217-236

\noindent Marcy, G.W. \etal 2001 http://exoplanets.org/index.html and http://exoplanets.org/science.html

\noindent Marcy, G.W. \& Butler, R.P. `2000 Millenium Essay: Planets 
Orbiting Other Suns' PASP 112, 137-140

\noindent Mayor, M. \etal 2001 Geneva Observatory 
http://obswww.unige.ch/~udry/planet

\noindent Naef, D. \etal 2001, ``The CORALIE survey for southern extrasolar 
planets V. 3 new extrasolar planets'' astro-ph/0196255

\noindent Papaloizou, J.C.B. \& Terquem, C. 1999, ``Critical protoplanetary 
core masses in protoplanetary disks and the formation of short period 
giant planets'' \apj, 521, 823-828

\noindent Tabachnik, S. \& Tremaine, S. 2001, ``Maximum-likelihood method for estimating
the mass and period distribution of extrasolar planets'' astro-ph/0107482

\noindent Tinney, C.G. \etal, 2002 
``Two Extra-solar Planets from the Anglo-Australian Planet Search'' 
\apj, submitted, astro-ph/0111255

\noindent Vogt, S. \etal 2002 
``Ten Low Mass Companions from the Keck Precision Velocity Survey''
\apj, submitted, astro-ph/0110378

\noindent Ward, W. 1997
``Survival of Planetary Systems'' \apj, 482, L211-L214

\noindent Weidenschilling, S.J. \& Marzari, F. 1996, ``Gravitational Scattering
as a possible origin for giant planets at small stellar distances''
Nature, 384, 619-621

\noindent Wetherill, G.W. 1994, ``Possible Consequence of absence of Jupiters in
planetary systems'' Ap\&SS, 212, 23

\noindent Wetherill, G.W. 1995, ``Planetary Science -- how Special is 
Jupiter?''  Nature, 373, 470

\noindent Zucker, S. \& Mazeh, T. 2001a, ``Derivation of the mass distribution of 
extrasolar Planets with MAXLIMA - a maximum likelihood Algorithm'' 
astro-ph/0106042  

\noindent Zucker, S. \& Mazeh, T. 2001b, ``Analysis of the Hipparcos Observations of
Extrasolar Planets and the Brown Dwarf Candidates''
astro-ph/0107142  

\noindent Zuckerman, B., Forveille, T. \& Kastner, J.H. 1995, `Inhibition of giant-planet formation by rapid gas 
depletion around young stars?'  Nature, 
373, 494-496 

\end{document}